\documentstyle[12pt,prd,eqsecnum,aps]{revtex}

\def\footnoterule{\kern-3pt \hrule width \hsize \kern2.6pt}
\newcommand{\beq}{\begin{equation}}
\newcommand{\eeq}{\end{equation}}
\def\A{{\bf A}}
\def\D{{\bf D}}

\def\bea{\begin{eqnarray}}
\def\eea{\end{eqnarray}}
\newcommand{\ba}{\begin{array}}
\newcommand{\ea}{\end{array}}
\def\nablar{\stackrel{\rightarrow}{\nabla}}
\def\nablaa{\stackrel{\leftrightarrow}{\nabla}}

\def\p{{\bf p}}
\def\x{{\bf x}}
\def\y{{\bf y}}
\def\z{{\bf z}}
\def\u{{\bf u}}
\def\v{{\bf v}}
\def\k{{\bf k}}

\def\f{{\bf f}}
\def\sd{M\frac{\partial}{\partial M}}
\def\M{\tilde{M}}

\begin{document}
\baselineskip=15.5pt plus1pt minus1pt

\draft
\title{ Differential Regularization of a Non-relativistic Anyon Model
\footnote{This research is supported in part by
D.O.E. contract DE-AC02-76ER03069,
N.S.F. Grant PHY/9206867 and by
CICYT (Spain) under Grant AEN90-0040.}}
\author{ D.~Z.~Freedman}
\address{ Department of Mathematics and
Center for Theoretical Physics, \\
Massachusetts Institute of Technology, Cambridge, MA 02139}
\author{G.~Lozano}
\address{International Centre for Theoretical Physics, \\
PO Box 586, I-34100, Trieste, Italy}
\author{N.~Rius}
\address{Center for Theoretical Physics, Laboratory for Nuclear Science,
and Department of Physics, \\
Massachusetts Institute of Technology, Cambridge, MA\ \ 02139}
\maketitle
\vspace{\fill}

\begin{center}
{\large ABSTRACT}
\end{center}

\begin{abstract}
Differential regularization is applied to a field theory of a non-relativistic
charged boson field $\phi$ with $\lambda (\phi {}^{*} \phi)^2$
self-interaction and coupling to a statistics-changing $U(1)$ Chern-Simons
gauge field.  Renormalized configuration-space amplitudes for all diagrams
contributing to the $\phi {}^{*} \phi {}^{*} \phi \phi$ 4-point function,
which is the only primitively divergent Green's function, are obtained up to
3-loop order.  The renormalization group equations are explicitly checked, and
the scheme dependence of the $\beta$-function is investigated.  If the
renormalization scheme is fixed to agree with a previous 1-loop calculation,
the 2- and 3-loop contributions to $\beta(\lambda,e)$ vanish, and
$\beta(\lambda,e)$ itself vanishes when the ``self-dual'' condition relating
$\lambda$ to the gauge coupling $e$ is imposed.
\end{abstract}

\vspace{\fill}

\hbox to \hsize{CTP \#2216 \hfil hep-th/9306117 \hfil June 1993}

\thispagestyle{empty}
\newpage

\setcounter{page}{1}
\section{Introduction}

The differential renormalization procedure \cite{fjl}
is a novel method
for perturbative calculations in quantum field theory in which
real space correlation functions are simultaneously regularized
and renormalized.  Neither explicit cutoff nor singular counter
terms are required, yet no ultraviolet divergences appear.  The
method has now been applied to many relativistic field theories,
and a systematic formulation \cite{lmv},
obtained with the use of BPHZ
topological analysis of Feynman diagrams, has recently been
given.

Field theories with non-relativistic kinematics are also of
physical interest, and we report in this paper on the application
of differential regularization to a theory in two spatial
dimensions which has been used to describe the anyon
Aharonov-Bohm and fractional quantum Hall effects.  The
particular theory \cite{jp} we study has evolved from
earlier work \cite{h,zhk}.
It describes a non-relativistic boson field with quartic
self-interaction and coupling to a 2+1 dimensional abelian gauge
field with Chern-Simons kinematics.

It was shown in \cite{jp} that the classical Lagrangian of the theory
is scale invariant, but one expects this symmetry to be broken by
renormalization effects at the quantum level.  These
scaling properties are similar to those of renormalizable
massless relativistic theories in 3+1 dimensions, so the theory
is a natural place to apply a new regularization method.  The
theory is also chiral so that dimensional regularization is
problematic.

The renormalization properties of the theory have been studied at
the 1-loop level \cite{g,bl} using a momentum cutoff
to handle ultraviolet divergences.  Scale symmetry breaking is
found for general values of the coupling constants, but scale
invariance is restored when the couplings are related by the
``self-duality'' condition which has special significance in the
classical theory \cite{jp}.

In this work we use the differential regularization method to
renormalize the divergent Green's functions of the theory through
3-loop order.  This method is based on the observation that bare
amplitudes which are primitively divergent in momentum space are
well defined for separated points in real $\vec{x}, t\/$ space,
although too singular at short distance to possess a Fourier
transform.  These amplitudes are then renormalized by the
following two-step procedure:
\begin{enumerate}
\item[(1)] Use differential identities (typically containing a new
  mass scale parameter $M\/$) to express singular bare amplitudes
  as derivatives of less singular Fourier transformable
  functions.

\item[(2)] Define the renormalized Fourier transform and other
  integrals of the originally singular amplitudes by formal
  partial integration of derivatives.
\end{enumerate}
Multi-loop amplitudes are then handled by inserting the
differential identities for their primitively divergent
subgraphs, manipulating differential operators according to step
(2) and using new differential identities as needed.  The
heuristic reason that this procedure is consistent is that the
ultraviolet singular surface terms which are neglected in rule
(2) correspond to singular counter terms which occur in
traditional regularization methods.  The operational test of
consistency is that the amplitudes obtained obey renormalization
group equations (RGE's) in the scale parameter $M\/$, with
$\beta\/$-functions and anomalous dimension defined through the
equations themselves.

New features occur when these ideas are implemented in a
non-relativistic field theory.  Because of the lack of symmetry
between space and time coordinates, differential identities of
rather different structure from those for relativistic theories
must be found.  Due to the combined non-relativistic and
Chern-Simons kinematics the gauge coupling is not renormalized,
so that there is only a $\beta\/$-function for the quartic
coupling.  Nevertheless this $\beta\/$-funtion is
scheme-dependent beyond one-loop order, as expected for a theory
with two or more couplings.  In differential renormalization,
scheme dependence appears because of the freedom to introduce
different mass scales $M\/$, $M'\/$ in some of the differential
identities used to regulate the overall divergence of graphs which
are not related by symmetry.  If this occurs in a given order of
perturbation theory the $\beta\/$-function typically depends on
$\log M / M'\/$ in the next order and beyond.  Our calculations show
that there are two scale parameters which appear at the one-loop
level and one more at the three-loop level.
The dependence of
the two- and three-loop $\beta\/$-function on the ratio of the one-loop
scale parameters is obtained.
If the mass ratio is fixed to make our one-loop result
agree with \cite{g,bl}, then the two- and three-loop $\beta\/$-function
vanishes.  The remaining one-loop contribution then vanishes when
the two couplings satisfy the ``self-duality'' condition of
\cite{jp}.
Thus previous results on the relation of ``self-duality'' to
scale invariance in the quantum theory have now been extended
through three-loop order.

Our approach to this problem has been purely field-theoretic; we
have simply obtained renormalized real space Green's functions
which obey RGE's.  We have not compared results with closely
related quantum mechanical studies of anyons \cite{mt},
although it appears that some difficulties of perturbative
quantum mechanical treatments can be resolved using
field-theoretic methods.  Finally, we can only express the hope
that our work can be applied to elucidate physical properties of
anyon systems.\footnote{%
During the writing phase of this work, references
\cite{vascos} came to
our attention through the bulletin board.  Three-loop
calculations of the finite $V\/$, $T\/$ partition function of the
theory at the self-dual point are presented in these papers.}

The Lagrangian of the theory under study is
\beq
L = \phi^*
\left(
i D_t + \frac{1}{2m} \D^2
\right) \phi
+ \alpha \epsilon_{ij}
\left(
\frac12 \partial_t A_i A_j - A_0 \partial_i A_j
\right)
- \lambda (\phi^* \phi)^2
\label{1.1}
\eeq
with covariant derivatives
\beq
\begin{array}{rcl}
  D_t & = & \partial_t + ie A_0\\
  \D & = & \nabla - ie \A
\end{array}
\label{1.2}
\eeq
In the Coulomb gauge the only non-vanishing components of the
gauge field propagator take the instantaneous
form~\footnote {We use the notation $x = (\x,t)$.}
\bea
\langle
A_i (x) A_0 ( 0 ) \rangle &=& i T_i(x) \nonumber \\
T_i(x) &=&
\frac{ \delta (t)}{2 \pi \alpha} \ \epsilon_{ij} \
\frac{\x_j}{\x^2}
\label{1.3}
\eea
as required by the physical role of $A_\mu\/$ as a statistical
gauge field without propagating degrees of freedom.  The
Schr\"odinger propagator is
\beq
\begin{array}{rcl}
  \langle T \phi ( x ) \phi^* ( 0 ) \rangle & = & i G (x )\\
  G (x) & = & - \displaystyle
                   \frac{m}{2 \pi t} \theta ( t )
  e^{ \frac{i m \x^2}{2 t}}
\end{array}
\label{1.4}
\eeq
It satisfies
\bea
\left( i \partial_t + \frac{\nabla^2}{2 m} \right)
G (x) & = & \delta ( t ) \delta ( \x ) \\
\displaystyle
\lim_{t \rightarrow O^+} G (x) & = & -i \delta (\x).
\label{in5}
\eea
Perturbation calculations may be performed \cite{bl} with the
propagators (\ref{1.3}), (\ref{1.4})
and the interaction vertices from (1.1).
The superficial degree of divergence for the different Green functions
can be calculated by taking into account that the scale dimension
of time is twice that of the space coordinate.

Many Feynman graphs vanish due to the special kinematics of the
theory.  There is no particle production so the number of
$\phi\/$-field lines is conserved in intermediate states of any
graph.  Many graphs with internal gauge field lines vanish when
one adds the contributions where attachments of $A_0\/$ and
$A_i\/$ are exchanged.  These features of the theory imply that
there are no quantum loop contributions to the propagators (1.3)
and (1.4), the $\phi^* \phi A\/$ vertex and the $\phi^* \phi A
A\/$ ``Compton'' amplitude.  This means that the
only superficially divergent Green's function is the $\phi^*
\phi^* \phi \phi\/$ 4-point function, and there are no anomalous
dimensions in the theory.
{}From the
4-point amplitude one can construct 6-, 8-, 10-point functions,
etc., using a skeleton expansion without encountering further
ultraviolet divergences.

In Section II below we confine our attention to the scalar
subtheory of (1.1).  We discuss the basic differential identity
for the 1-loop scalar bubble graph, and its Fourier transform.
The convolution theorem can then be used to compute and sum
multiple bubbles and obtain the full off-shell scattering
amplitude.  In Section III, the regularization of the 1-loop
graphs of the full theory is obtained.  Two- and three-loop
contributions are considered in Sections IV and V, respectively.
The verification of the RGE's and the
$\beta\/$-function are discussed in Section VI.

\section{Scalar Sector}
\setcounter{equation}{0}

In this section we will study the scalar sector of the theory
defined by the Lagrangian (1.1).  The tree level contribution to
the amputated 4-point function is
\begin{equation}
  \Gamma^{a} (x_i) =
  -4i \lambda \delta (x_1-x_2) \delta (x_1-x_3) \delta (x_1-x_4)
  \label{s1}
\end{equation}
where we have used the simplified notation
$\delta ( x ) \equiv \delta ( t ) \delta (\x)\/$.

In a non-relativistic theory, the only non-vanishing one-loop
diagram is the direct channel bubble graph depicted in Fig.\ 1c.
The bare amplitude in configuration space is given by
\begin{equation}
  \Gamma^{c} ( x_i) = 8 \lambda^2
\delta (x_1-x_2) \delta (x_3-x_4)
G^2 ( x_3-x_1 )
  \label{s2}
\end{equation}
where $G( x_3-x_1 )\/$ is the scalar propagator given by eq.\
(1.4).  This amplitude is well-defined for separated points, but
the factor $1/ t^2\/$ is too singular at short distances,
yielding a logarithmically divergent Fourier transform.

The bare amplitude can be regulated by the following differential
identity
\begin{equation}
  \begin{array}{rcl}
    G^2_{reg}(x) & = &
      \displaystyle
      - \frac{im^2}{4\pi^2}
      \left(
      i \partial_t + \frac{\nabla^2}{4m}
      \right) F ( x ) \\
    F ( x ) & = &
      \displaystyle
      \log i M^2 t \frac{\theta ( t )}{t} e^{\frac{im \x^2}{t}}
  \end{array}
\label{s3}
\end{equation}
which is an exact relation away from the origin constructed to
incorporate the ideas discussed in the introduction.

i)  The differential operator acts on the function $F ( x )\/$,
which is less singular than $G^2 ( x )\/$ and has a well-defined
Fourier transform. Using Gaussian integration it is
straightforward to obtain
\begin{equation}
  \hat F ( \omega, \k ) =
  -\frac{\pi}{m}
  \frac{\log ( \k^2/4m - \omega - i \epsilon ) \gamma / M^2 ]}
       {\k^2/4m - \omega - i \epsilon},
\end{equation}
where $\gamma = 1.781\ldots\/$ is  Euler's constant.

ii)  The coincident point singularity of (2.2) is defined as in
distribution theory using formal partial integration of
derivatives in any integral.  For the Fourier transform, this
prescription gives
\begin{equation}
  \hat \Gamma^c (\omega, \k) =
  -\frac{2im\lambda^2}{\pi} \log
\left[ \left(\frac{\k^2}{4m}-\omega-i \epsilon \right)
{\gamma \over M^2} \right]
\label{s5}
\end{equation}

iii) Most important is the fact that the parameter $M\/$, which
is required for dimensional reasons in (\ref{s3}), plays the role of
the renormalization group scale parameter.  To see this we can
compute the effect of a variation of $M\/$
\begin{equation}
  \renewcommand{\arraystretch}{2}
  \begin{array}{rcl}
    \displaystyle
    M \frac{\partial}{\partial M} G^2_{reg}(x)& = &
    \displaystyle
- \frac{im^2}{2\pi^2} \left( i \partial_t + {\nabla^2 \over 4m}\right)
    \frac{\theta ( t )}{t} e^{\frac{im \x^2}{t}}\\
    & = & \displaystyle
    \frac{i m}{2 \pi} \delta (x)
  \end{array}
\label{s6}
\end{equation}
where we have used (\ref{in5}) and the fact that the $\log i M^2 t\/$
factor in (2.3) multiplies a Schr\"{o}dinger propagator with
mass $2m\/$.  Comparing (\ref{s1}, \ref{s2}, \ref{s6}),
we see that the
effect of a change in $M\/$ can be compensated by a change in the
coupling constant.  Indeed from the renormalization group
equation
\begin{equation}
  M \frac{\partial}{\partial M} \Gamma_{(s)} ( x_i ) =
  - \beta ( \lambda ) \frac{\partial}{\partial\lambda}
  \Gamma_{(s)} ( x_i )
\end{equation}
with $\Gamma_{(s)}\/$ taken as the sum of the tree and
renormalized bubble amplitudes one finds the one-loop
$\beta\/$-function
\begin{equation}
  \beta ( \lambda ) = \frac{m \lambda^2}{\pi}
\label{s8}
\end{equation}

iv)  The factor of $i\/$ in $\log i M^2 t\/$ is
required to obtain a real analytic amplitude in momentum space
with non-zero imaginary part for $\omega > \k^2/4m\/$ as required
by unitarity.  Alternatively, as in the relativistic case [1],
one could start in the imaginary time formalism, $t = -i \tau\/$,
in which all bare amplitudes and the differential identities used
to regulate them are real functions of $x\/$ and $\tau\/$.  Then
$\log i M^2 t\/$ is automatically obtained for real
Schr\"{o}dinger time.

In the scalar theory it is easy to sum multi-bubble graphs and
obtain the full scattering amplitude.  These graphs are
convolutions in real-space and the renormalized amplitudes in
$p\/$-space can be defined as products of the renormalized
1-bubble amplitude (2.5).
The scattering amplitude is then
\begin{equation}
  A ( \omega, \k ) = - 4 i \lambda
  \sum^\infty_{n = 0}
  \left(
  \frac{i}{4\lambda} \hat \Gamma^c ( \omega, \k )
  \right)^n.
\end{equation}
In the c.m.\ frame, $\k = 0\/$, one obtains
\begin{equation}
  A ( \omega, 0 ) = -
  \frac{4 i \lambda}
{1 - \displaystyle \frac{m \lambda}{2\pi} \log
  \frac{-(\omega + i \epsilon) \gamma}{M^2}}.
\end{equation}

The above remarks are intended to show how the differential
renormalization procedure reproduces previous results on the
scalar sector \cite{j,b}, obtained in most cases with a momentum
cutoff and renormalization at a momentum scale $\p^2 = \mu^2\/$
(here $\p=\p_1=-\p_2$, where $\p_1$ and $\p_2$ are the
momenta of the incident particles). Since the on-shell c.m.\ total energy
is given by
$\omega = \p^2/m\/$,
our renormalized amplitude can be brought
into full agreement with previous work if the scale parameters
are related by $\mu^2 = m M^2 / \gamma\/$.

\section{One loop order}
\setcounter{equation}{0}

In this section we shall illustrate how differential regularization
works for the full theory to one-loop order.
The tree graph of Fig. 1b does not require regularization, but for
completeness we give its $x$-space amplitude,
\beq
\Gamma^b (x_i) = \frac{e^2}{m}\delta(x_1-x_3)T_i(x_1-x_2)
\frac{\partial}{\partial\x_4^i}\delta(x_4-x_2)
\eeq

The one-loop graphs are shown in Fig. 1c-1f. In addition to the
diagrams explicitly drawn one must add, as appropriate in each
case, i)the exchange of $A_0$ and $A_i$ vertices,
ii)the opposite orientation of each triangle with $A_0, A_0, A_i^2$
vertices, iii)the time-reflected graph with initial and
final states interchanged, iv)exchange of $x_3$ and $x_4$ as required
by Bose symmetry. It would be tedious to discuss these permutations
explicitly, but it is important to include them when the
RGE's are discussed in section VI.

Power counting leads us to expect that all one-loop graphs for
the 4-point function are ultraviolet divergent. However,
as we shall show, the only 1-loop graph which actually requires
renormalization (apart from the bubble already discussed in section
II) is the seagull graph shown in Fig. 1d. We therefore begin
with this graph whose bare amplitude is given by the following
expression,

\beq
\Gamma^{d}(x_i)= \frac{e^4}{m} \delta(x_2-x_4) G(x_3-x_1)
T_i(x_2-x_1) T_i (x_2-x_3).
\label{m2}
\eeq
Note that in this expression there is a factor
\beq
\theta(t_3-t_1)\delta(t_3-t_1)
\eeq
where the $\theta$  and $\delta$ function come from the boson and
gauge field propagator respectively. We then use that,
\beq
\theta(t)\delta(t)=\frac{1}{2}\delta(t)
\eeq
a property that can be deduced from the convolution theorem.
Using  this together with the value at $t=0$ of the boson propagator
(\ref{in5}), we obtain,
\beq
\Gamma^d(x_i) = -\frac{ i e^4}{8m \pi^2 \alpha^2} \delta(x_2-x_4)
                     \delta(x_1-x_3)\delta(t_3-t_2) \frac{1}{(\x_2-\x_1)^2}
\label{3.8}
\eeq

Notice that as the boson propagator at $t=0$ is proportional
to a delta function, the triangle ``collapses" to an effective
gauge field bubble.
The function $\frac{1}{(\x_1-\x_2)^2}$ is too singular at short distances
and as a result, its Fourier transform is logarithmically divergent.
In the spirit of differential regularization, we use the following
identity, valid at all points except $\x=0$,
\beq
\frac{1}{\x^2}= \frac{1}{8} \nabla^2 \log^2M_s^2\x^2
\label{3.9}
\eeq

Note that we have introduced a different mass scale from that in
the identity (\ref{s3}) associated with the bubble graph. A priori, one
can use independent scales for different graphs which are not related by
a Ward identity. Different choices
of M's correspond to different renormalization schemes and in
section VI we shall
see how this affects the $\beta$-function.

After using equation (\ref{3.9}), we obtain the regularized form
\beq
\Gamma^d_{reg}
(x_i) = -\frac{ i e^4}{64m \pi^2 \alpha^2} \delta(x_2-x_4)
                     \delta(x_1-x_3)\delta(t_3-t_2) \nabla^2_{_1} \log^2
M_s^2 (\x_1-\x_2)^2
\label{3.10}
\eeq
which has now a well defined Fourier transform, since it can be shown
using the same technique as in the Appendix of \cite{fjl} that
\beq
\int d^2x e^{i\p \x} \log^2M^2\x^2=-\frac{8\pi}{\p^2}\log\frac{4M^2}{\gamma^2
\p^2}
\eeq
{}From this expression we can calculate the contribution of
this graph to the 4-point function in momentum space.
After taking into account the graphs obtained by permutation, we have
in the center of mass frame,
\beq
\hat \Gamma^D (\p)=
-\frac{i e^4}{2\pi \alpha^2} \log\left(\frac{\M_s^2}{2\p^2
|\sin\theta|}\right)
\label{3.12}
\eeq
which coincides with the result in ref. \cite{bl} if the identification
$\mu=\M_s=\frac{2M_s}{\gamma}$ is made.

Comparing (\ref{3.12}) and the momentum space amplitude of the
bubble graph (\ref{s5}) in the c.m. frame
we also see that the renormalization scheme adopted in \cite{bl}
corresponds to the following relation between scales
\beq
M_s^2 = \rho_1 M^2= \frac{m\gamma}{4}M^2
\label{3.13}
\eeq

We end this section by showing that the remaining one-loop graphs
do not require renormalization.
It can be easily shown that contribution of the triangle graph shown in
Fig. 1e cancels against the one coming from the graph with
the $<A_0A_i>$ propagator running in the opposite direction.
Of course, any higher-loop diagram that
contains this one as subdiagram will automatically
vanish, therefore they are not included in Fig. 1
and we shall omit them from our discussion.

We shall show next that contributions of the box graphs are finite.
Let us consider
as an example the diagram shown in Fig. 1f,  which is given by
\beq
\Gamma^{f}(x_i)=- \frac{e^4}{m^2} G(x_3-x_1)
\left(\frac{\partial}{\partial \x^i_2}\frac{\partial}{\partial \x_4^j}
G(x_4-x_2)\right) T_i(x_2-x_1)T_j(x_4-x_3)
\label{3.1}
\eeq
Because the gauge field propagator is transverse,
\beq
\nabla_{\x} \cdot {\bf T}(x-y)=0
\label{3.2}
\eeq
one can immediately extract the derivatives as total derivatives and
rewrite the amplitude as,
\beq
\Gamma^{f}(x_i)=- \frac{e^4}{m^2} \frac{\partial}{\partial \x_2^i}
\frac{\partial}{\partial \x_4^j}
 [G(x_3-x_1)G(x_4-x_2) T_i(x_2-x_1)T_j(x_4-x_3)]
\eeq

The Fourier transform of this amplitude can be expressed now
as a finite loop momentum integral and the total derivatives
are simply factors of external momenta. The other box graphs
with $A_o\leftrightarrow A_i$ permutations can be treated similarly.

\section{Two loop order}
\setcounter{equation}{0}

At two-loop order there are two  divergent graphs: the double bubble shown
in Fig. 1g and the ``ice cream" graph shown in Fig. 1h.

A regularized expression for the double bubble is obtained immediately
using the regulated expression (\ref{s3}) for each bubble subgraph,
\beq
\Gamma^{g}_{reg}(x_i)=
16i\lambda^3 \delta(x_1-x_2)\delta(x_3-x_4)
\int d^3 u G^2_{reg}(x_3-u)G^2_{reg}(u-x_1)
\label{4.1}
\eeq
This integral can be explicitly computed and we obtain
\beq
\Gamma^{g}_{reg}(x_i)=i
\left( \frac{m\lambda}{\pi}\right)^3
\delta(x_1-x_2)\delta(x_3-x_4)
\left(i\partial_{t_3} + \frac{\nabla^2_3}{4m}\right)S(x_3-x_1)
\label{4.2}
\eeq
where
\beq
S(x)=\frac{\theta(t)}{t}e^{\frac{im\x^2}{t}}\left(\log^2iM^2t-\frac{\pi^2}{6}
\right)
\label{4.3}
\eeq
Now, let us turn to the regularization of the ice cream graph.
The bare amplitude is given by
\beq
\Gamma^{h}(x_i)=\frac{e^4 \lambda}{2m\pi^2\alpha^2}\delta(x_1-x_2)
\delta(t_3-t_4)G(x_3-x_2)G(x_4-x_1) \frac{1}{(\x_3 -\x_4)^2}
\label{4.4}
\eeq
For simplicity, let us set $x_1=0$ and introduce coordinates $\x$, $\y$
and $t$ such
that $x_3=(\x,t)$ and $x_4=(\y,t)$. The singular part of the graph is then
contained in
\beq
f^{h}(\x,\y,t)=  G(\x,t) G(\y,t)\frac{1}{(\x-\y)^2}.
\label{4.5}
\eeq
We begin  by regularizing the divergent seagull subgraph
(see Eq. (\ref{3.9}))
\beq
f^{h}(\x,\y,t) \rightarrow f^{h}(\x,\y,t)=\frac{1}{32} G(\x,t) G(\y,t)
(\nabla_x -\nabla_y)^2 \log^2M_s^2 (\x-\y)^2
\eeq
After manipulating derivatives, this expression can be rewritten as
\bea
f^{h}(\x,\y,t) &=&\frac{1}{32}
\left\{-(\nablar_\x-\nablar_\y)\left[\log^2(M_s^2
(\x-\y)^2)(\nablaa_\x -\nablaa_\y)G(\x,t) G(\y,t)\right] \right. \nonumber \\
 &  & \left. + \log^2 M_s^2(\x-\y)^2
(\nablar_x-\nablar_y)^2 G(\x,t) G(\y,t) \right\}
\label{4.7}
\eea
The first term in this expression, being a total divergence, is
finite by power counting. In order to improve the second term,
note that the product of two non-relativistic propagators satisfies
the identity
\beq
(\nablar_x-\nablar_y)^2 G(\x,t) G(\y,t) = - \left[ (\nablar_x +\nablar_y)^2
+ 4im  \partial_t \right] G(\x,t) G(\y,t) - 4im\delta(\x)\delta(\y)\delta
(t)
\label{4.8}
\eeq
To regularize (\ref{4.7}) we drop the last term in this identity,
because it simply gives a singular surface term in the Fourier
transform of $f^h$.
In this way, we obtain the regularized expression
\bea
f^{h}_{reg}(\x,\y,t)&=&-\frac{1}{32}
\left\{(\nablar_\x-\nablar_\y)\left[\log^2M_s^2(\x-\y
)^2 (\nablaa_\x -\nablaa_\y)G(\x,t) G(\y,t)\right] \right. \nonumber \\
 & & \left. + \left[ (\nablar_\x +\nablar_\y)^2  + 4im  \partial_t \right]
\log^2M_s^2(\x-\y)^2 G(\x,t) G(\y,t)\right\}
\label{4.9a}
\eea

This can be inserted in (\ref{4.4}), restoring the original
space-time points,
to obtain a regularized form for the amplitude $\Gamma^h(x_i)$.

We shall show next that the rest of the two-loop graphs in Fig. 1 are
indeed finite.
The bare amplitude for diagram i is
\beq
\Gamma^{i}(x_i)=-\frac{ie^6}{24\pi^2\alpha^2}\left[ \frac{\partial}
{\partial \x_2^i}G(x_4-x_2)\right]G(x_3-x_1)T_i(x_2-x_1)
\frac{1}{(\x_4-\x_3)^2}
\eeq
We first regularize the subdivergence $\x_3 \rightarrow \x_4$
by using the identity (\ref{3.9})
\beq
\Gamma^{i}(x_i)=-\frac{ie^6}{24\pi^2\alpha^2}
\left[ \frac{\partial}{\partial \x_2^i}
G(x_4-x_2)\right] G(x_3-x_1)T_i(x_2-x_1)
\frac{1}{8}\nabla^2 \log^2M_s^2(\x_4-\x_3)^2
\eeq

Using now the property (\ref{3.2}), the overall divergence can
be regularized by writing the diagram as total derivative,
\beq
\Gamma^{i}(x_i)=-\frac{ie^6}{24\pi^2\alpha^2}\frac{\partial}
{\partial \x_2^i}\left[G(x_4-x_2)G(x_3-x_1)T_i(x_2-x_1)
\frac{1}{8}\nabla^2 \log^2 M_s^2 {(\x_4-\x_3)^2}\right]
\label{4.9}
\eeq

Diagram i is now regularized. Nevertheless, let us point
the following fact. Though this diagram depends on $M_s$, the sum
of this diagram plus the one where the $<A_0A_i>$ line runs
in the opposite direction {\it does not}. This can be understood
in the following way: as easily seen from (\ref{4.9}), the scale derivative
of $\Gamma^{i}$ is proportional to $\Gamma^e$, and we have
already shown that this diagram is zero once the corresponding
$A_0 \leftrightarrow A_i$ exchange is taken into account. We then see
that although a scale dependence is introduced in intermediate
steps to properly handle subdivergences, this dependence can
disappear in the final amplitude.
As we shall see, a similar situation occurs in several three-loop
diagrams.

The last two-loop diagram, Fig. 1j, can be shown to be finite.
After some manipulation of derivatives, the amplitude can
be written as
\beq
\Gamma^j(x_i)=-\frac{e^6}{m^3}\frac{\partial}{\partial \x_2^i}
\frac{\partial}{\partial \x_4^k} T_i(x_2-x_1)T_k(x_4-x_3)
\frac{\partial}{\partial \x_4^j}\int_{u,v} G(x_3-v)G(x_4-u)
G(u-x_2)G(v-x_1)T_j(u-v)
\eeq
The Fourier transform of this amplitude is a two-loop
energy-momentum integral which is convergent by power counting because
it contains explicit factors of external momenta.

\section{Three loop order}
\setcounter{equation}{0}

There are eight diagrams contributing to the 4-point function
at three loops (see Fig. 1)
although we shall show that only four of
them are divergent, namely diagrams k-n.
To regularize three-loop bare amplitudes we first identify the
divergent subgraphs and treat them in the same way as we have
done at lower orders. The regularization of overall divergences
(when all the external points come close) may require
new DR identities.
Our main concern is to compute the contribution of every diagram
to the RGE's
(see section VI) and thus
in some cases we won't give the explicit closed form of the diagram,
since it is not needed,
but only an integral representation that has an ultraviolet
finite Fourier transform and from which we can obtain the relevant
contributions of the diagram to the Callan-Symanzik equations.

Let us start with the triple bubble diagram. We regulate each bubble
subgraph using eq. (\ref{s3}) and we obtain
\beq
\Gamma_{reg}^k (x_i) =
- 32 \lambda^4 \delta(x_1-x_2) \delta(x_3-x_4)
\int_{u,v} G^2_{reg}(u-x_1) G^2_{reg}(v-u) G^2_{reg}(x_3-v)
\label{b1}
\eeq
This expression is the convolution of three factors which have a finite
Fourier transform, therefore so does the full result, and we shall
find the contribution to the
RGE's explicitly from (\ref{b1}).

Diagram l has also the structure of a convolution of two
subdiagrams we have already regulated, namely the one bubble graph
(\ref{s3}) and the ``ice cream cone'' (\ref{4.9a}).
Using these we obtain
\bea
\Gamma_{reg}^l(x_i) &=&
{ i \lambda^2 e^4 \over m (\pi \alpha)^2}
\delta(x_1-x_2) \delta(t_3-t_4)
\nonumber \\
&\times&
\int dt d^2\u G^2_{reg} (u-x_1; M)
f_{reg}^h(\x_3-\u, \x_4-\u, t_3-t; M_s)
\label{bi1}
\eea
where, as we have explicitly indicated, the
mass scales used in regulating the subdivergences are different.
Again, (\ref{bi1}) is an adequate regulated expression,
since each factor in the convolution has a well defined Fourier
transform, and thus it is all we need to
find the contribution of the diagram to the Callan-Symanzik equations.

The regularization of diagram m is more involved.
The bare amplitude is
\beq
\Gamma^m(x_i)=
{1 \over 64 m^2 }
\left({e^2 \over \pi\alpha} \right)^4
\delta(t_2-t_1) \delta(t_4-t_3) f^m(x_i)
\label{t1}
\eeq
where
\beq
f^m(x_i) =
{1 \over (\x_2-\x_1)^2 (\x_4-\x_3)^2}
G(x_3-x_1) G(x_4-x_2)
\label{t2}
\eeq
After relabelling points
$\x_1 \rightarrow \x$,
$\x_2 \rightarrow \y$,
$\x_3 \rightarrow \z$,
$\x_4 \rightarrow 0$,
$t_1=t_2 \rightarrow 0$, $t_3=t_4 \rightarrow t$ and
regulating the two divergent seagull subgraphs by using (\ref{3.9}),
we obtain the partially regulated expression:
\beq
f^m(\x,\y,\z,t)=
{1\over 16^2}
G(\z-\x,t) G(-\y,t)
(\nabla_\x - \nabla_\y)^2 \log^2 M_s^2 (\x-\y)^2
\nabla^2_\z \log^2 M_s^2 \z^2
\label{t3}
\eeq
Now we integrate by parts the differential operator
$(\nabla_\x - \nabla_\y)^2$ and we use the relation
(\ref{4.8}) satisfied by the product of two scalar propagators.
Note that when we insert the exact identity (\ref{4.8}) in (\ref{t3})
we have to keep the term that contains the
$\delta$-functions because
it is not a purely local contribution
(as it was in the ``ice cream cone'' amplitude (\ref{4.8}), where we
dropped it),
but it multiplies a factor which requires further regularization.

We then rewrite (\ref{t3}) as
\bea
f^m(\x,\y,\z,t) &=&
-{1\over 16^2}
\left\{(\nablar_\x-\nablar_\y)\left[ \log^2 M_s^2(\x-\y)^2
(\nablaa_\x -\nablaa_\y)G(\z-\x,t) G(-\y,t)
\nabla^2_\z \log^2 M_s^2 \z^2 \right] \right. \nonumber \\
 &  &
\left. + \left[ (\nablar_\x + \nablar_\y)^2 + 4 i m \partial_t \right]
\log^2 M_s^2(\x-\y)^2 G(\z-\x,t) G(-\y,t)
\nabla^2_\z \log^2 M_s^2 \z^2 \right. \nonumber \\
& &
\left. + 32 i m \delta(t) \delta(\z-\x) \delta(\y)
{\log^2 M_s^2\x^2 \over \x^2}
\right\}
\label{t5}
\eea
Note that once we have regulated the
subdivergence $\z \sim 0$ by using (\ref{3.9}),
we do not need to integrate by parts the operator $\nabla^2_\z$
as we did with $(\nabla_\x - \nabla_\y)^2$.
The reason is that the first two terms are
already total derivatives with respect to external points and,
since the diagram is only logarithmically divergent, these terms have a
finite Fourier transform, defined by partial integration neglecting
surface terms.
The remaining term is regulated using the new DR identity
\beq
{\log^2 M_s^2 \x^2  \over \x^2} =
{1 \over 48} \nabla^2 \log^4 M_s^2 \x^2
\label{t6}
\eeq
Thus the final regulated amplitude is
\bea
\Gamma^m_{reg}(x_i) &=&
- \left({e^4 \over 2 m} \right)^2
{1 \over (8\pi\alpha)^4}
\delta(t_2-t_1) \delta(t_4-t_3)  \nonumber \\
&\times &
\left\{(\nablar_1-\nablar_2)\left[ \log^2 M_s^2(\x_1-\x_2)^2
(\nablaa_1 -\nablaa_2)G(x_3-x_1) G(x_4-x_2)
\nabla^2 \log^2 M_s^2 (\x_3-\x_4)^2 \right] \right. \nonumber \\
 &  &
\left. + \left[ (\nablar_1 + \nablar_2)^2 + 4 i m \partial_{t_3} \right]
\log^2 M_s^2(\x_1-\x_2)^2 G(x_3-x_1) G(x_4-x_2)
\nabla^2 \log^2 M_s^2 (x_3-x_4)^2 \right. \nonumber \\
& &
\left. + {2 i m \over 3} \delta(t_3-t_1)
\delta(\x_1-\x_3) \delta(\x_2-\x_4))
\nabla^2 \log^4 M_s^2(\x_1-\x_2)^2
\right\}
\label{t7}
\eea

The bare amplitude of diagram n is given by
\beq
\Gamma^n(x_i) =
{2i \over m}
\left( {\lambda e^2 \over \pi\alpha} \right)^2
\delta(x_1-x_2) \delta(x_3-x_4) I(x_3-x_1)
\label{c1}
\eeq
where
\beq
I(x_3-x_1)=
\int_{u,v} G(u-x_1) G(v-x_1) G(x_3-u) G(x_3-v) \delta(t_u-t_v)
{1 \over (\u-\v)^2}
\label{c2}
\eeq
This is the first diagram which involves an internal integration which
is not a convolution. Indeed this integral contains a
logarithmic subdivergence for $\u \sim \v$.
Our strategy is to regulate the singular factor
$1/(\u-\v)^2$ using (\ref{3.9}) and then to
perform the integral over internal points using formal partial integration.
After this is done we will cure the overall divergence when
$x_3 \sim x_1$.

We relabel points as $x_1 \rightarrow 0$, $x_3 \rightarrow (\x,t)$,
$t_u = t_v \rightarrow t'$.
After regulating the divergent seagull subgraph by using (\ref{3.9})
we integrate by parts the laplacian operators.
Total derivatives with respect to internal points can be dropped and
we obtain the partially regulated form
\bea
I(\x,t) &=& {1 \over 32}
\int dt'd^2\u d^2\v
\log^2 M_s^2(\u-\v)^2
\label{c3} \\
&\times &
\left\{
\left[(\nablar_\u - \nablar_\v)^2 G(\u,t') G(\v,t') \right]
G(\x-\u,t-t') G(\x-\v,t-t') \right.  \nonumber \\
& &
+ \left.
G(\u,t') G(\v,t')
\left[(\nablar_\u - \nablar_\v)^2 G(\x-\u,t-t') G(\x-\v,t-t')\right]
 \right.  \nonumber \\
& &
+ 2 \left.
\left[ (\nablar_\u - \nablar_\v) G(\u,t') G(\v,t') \right] \cdot
\left[ (\nablar_\u - \nablar_\v) G(\x-\u,t-t') G(\x-\v,t-t') \right]
\right\}  \nonumber
\eea
The first term, which we shall denote $I_1(\x,t)$,
can be easily regulated by means of the DR identity
(\ref{4.8}).
The delta term can be dropped because its effect is cancelled by the
counterterm which cancels the corresponding singular surface term
in the ice cream cone subgraph.
After partially integrating the differential operator and
replacing derivatives with respect to internal points
by derivatives with respect to $\x$, $t$,
we can rewrite $I_1(\x,t)$ as
\bea
I_1(\x,t) &=& - {1 \over 32}
(4 i m \partial_t + \nabla^2_\x)
\label{c4} \\
&\times&
\int dt' d^2\u d^2\v
\log^2 M_s^2(\u-\v)^2
G(\u,t') G(\v,t') G(\x-\u,t-t') G(\x-\v,t-t')
\nonumber
\eea
The integral in (\ref{c4}) is now finite by power counting and
it can be done using Gaussian integration,
leading to the regulated result:
\bea
I_1(\x,t) &=& {m^2 \over 128 \pi^2}
(4 i m \partial_t + \nabla^2_\x)
\left\{
{\theta(t) \over t}
e^{{i m \x^2 \over t}}
\right. \nonumber \\
& &
\times  \left. \left[
\log^2 iM_s^2t
- 2 (\log \rho_1 + 2) \log iM_s^2t + k
\right] \right\}
\label{c5}
\eea
with
$k= \log^2\rho_1 + 4 \log\rho_1 + 8 - \pi^2/6$
and $\rho_1 = m \gamma/4$.

Note that (\ref{c5}) has a finite Fourier transform,
since $I_1(\x,t)$ was logarithmically divergent and it contains
total derivatives with respect to external points.
It can be shown that the second integral in (\ref{c3}) is equal to
$I_1(\x,t)$, so we now focus on the last and most involved one
\beq
I_3(\x,t) = - {m^2 \over 16}
\int dt' d^2\u d^2\v
{(\u-\v)^2 \over t' (t-t')}
\log^2 M_s^2(\u-\v)^2
G(\u,t') G(\v,t') G(\x-\u,t-t') G(\x-\v,t-t')
\label{c6}
\eeq
It is convenient to rewrite
\beq
{1 \over t' (t-t')} = {1 \over t}
\left( {1 \over t'} + {1 \over t-t'} \right)
\label{c7}
\eeq
so $I_3(\x,t)$ becomes the sum of
two terms, which can be shown to be equal using a change of the
integration variables.
Then we use the identity
\beq
m^2 (\u-\v)^2
{\theta(t-t') \over (t-t')^3}
e^{{i m \over 2(t-t')} [(\u-\x)^2 + (\v-\x)^2]} =
(4 i m \partial_t + \nabla^2_\x)
{\theta(t-t')  \over t-t'}
e^{{i m \over 2(t-t')} [(\u-\x)^2 + (\v-\x)^2]}
\label{c8}
\eeq
to obtain
\bea
I_3(\x,t) &=& - \left({m \over 2 \pi} \right)^4
{1 \over 8 t}
(4 i m \partial_t + \nabla^2_\x) \theta(t) \nonumber \\
&\times&
\int_0^t dt'\int d^2\u d^2\v
\log^2 M_s^2(\u-\v)^2
{e^{{i m \over 2t'} (\u^2 + \v^2)}  \over t'^2}
{e^{{i m \over 2(t-t')} [(\u-\x)^2 + (\v-\x)^2]} \over t-t'}
\label{c9}
\eea
We can easily perform the integral over internal points
and we find the partially regulated form
\bea
I_3(\x,t) &=& {m^2 \over 32 \pi^2 t}
(4 i m \partial_t + \nabla^2_\x)
\left\{ \theta(t)
e^{{i m \x^2 \over t}}  \right. \nonumber \\
&\times  &
\left. \left[
{1 \over 2} \log^2 iM_s^2t
- (\log \rho_1 + 2) \log iM_s^2t
\right. \right.    \nonumber \\
& & \left. \left.
+\left({1 \over 2}\log^2\rho_1 + 2 \log\rho_1 + 4 -{\pi^2 \over 12} \right)
\right] \right\}
\label{c10}
\eea
which still contains an overall divergence when $\x \sim 0$, $t \sim 0$.
To regulate it, we integrate by parts the time derivative so that we
are left with a term which is a total external derivative and therefore
has a finite Fourier transform, plus three remaining singular terms
that can be easily regulated by using (\ref{s3}) and two new DR
identities, namely
\beq
{\theta(t) \over t^2}  e^{{i m \x^2 \over t}}
\log^n iM_s^2t
=
-{i \over n+1}
\left(i \partial_t + {\nabla^2_\x \over 4 m} \right)
{\theta(t) \over t}  e^{{i m \x^2 \over t}}
\log^{n+1} iM_s^2t   \hspace{.5in} n=1,2
\label{c11}
\eeq
The final regulated form of $I_3(\x,t)$ is then
\bea
I_3(\x,t) &=& {m^2 \over 32 \pi^2}
(4 i m \partial_t + \nabla^2_\x)
\left\{ {\theta(t) \over t}
e^{{i m \x^2 \over t}} \right. \nonumber \\
&\times  &
\left. \left[
{1 \over 6} \log^3 iM_s^2t
- {1 \over 2} (\log \rho_1 + 1) \log^2 iM_s^2t
\right. \right.    \nonumber \\
& & \left. \left.
+\left({1 \over 2}\log^2\rho_1 + \log\rho_1 + 2 - {\pi^2 \over 12} \right)
\log iM_c^2t  \right] \right\}
\label{c12}
\eea
Note the presence of a new mass scale $M_c$
in (\ref{c12}), which appears
when regularizing the last term in (\ref{c10}).
It is easily seen that the choice of this mass parameter does not
modify the $\beta$-function at the three-loop level, so we
choose its value such that it cancels the local term
(last one) in (\ref{c5}) and it simplifies the expression of the
renormalized amplitude.
Restoring the original space-time points we obtain
\bea
\Gamma^n_{reg}(x_i) &=&
{ i \over 8}
\left( {m \lambda e^2 \over \pi^2 \alpha } \right)^2
\delta(x_1-x_2) \delta(x_3-x_4)
\left(i\partial_{t_3} + {\nabla^2_3 \over 4m} \right)
\left\{
{\theta(t_3-t_1) \over t_3-t_1}
e^{{i m (\x_3-\x_1)^2 \over t_3-t_1}}
\right. \nonumber \\
&\times&
\left. \left[
{1 \over 3} \log^3 iM_s^2(t_3-t_1)
- \log \rho_1 \log^2 iM_s^2(t_3-t_1)
\right. \right.    \nonumber \\
& & \left. \left.
+\left(\log^2\rho_1 - {\pi^2 \over 6} \right)
\log iM_s^2(t_3-t_1) \right] \right\}
\label{c13}
\eea
We will show now that the remaining diagrams of Fig. 1 although
superficially log divergent are instead finite.
The contributions of diagrams $o,p,q$ are given by,
\bea
\Gamma_{reg}^o(x_i) &=&
{i \over 32}
\left({e^4 \over m \pi \alpha} \right)^2
{\partial \over \partial \x_4^j}
\left \{ T_j(x_4-x_3)
\delta(t_1-t_2)  \nabla_1^2 \log^2 M_s^2 (\x_1-\x_2)^2
\right.
\nonumber \\
&\times & \left.
{\partial \over \partial \x_2^i}
\int_{u,v} G(u-x_1) G(v-x_2) G(x_3-u) G(x_4-v) T_i(v-u) \right\} \\
\Gamma^p_{reg}(x_i) &=&
{i \over 32}
\left({e^4 \over m \pi \alpha} \right)^2
{\partial \over \partial \x_2^i}
{\partial \over \partial \x_4^j}
\left \{
 T_i(x_2-x_1) T_j(x_4-x_3)  (\nabla_1 + \nabla_3)^2 \right.
\nonumber \\
&\times & \left.
\int dt d^2\u d^2\v \log^2 M_s^2(\u-\v)^2
G(u-x_1) G(v-x_2) G(x_3-u) G(x_4-v) \right\} \\
\Gamma^q_{reg}(x_i) &=&
- {\lambda \over 16}
\left({e^3 \over m \pi \alpha} \right)^2
\delta(x_1-x_2)
{\partial \over \partial \x_4^j}
\left \{
 T_j(x_4-x_3)  (\nabla_1 + \nabla_3)^2 \right.
\nonumber \\
&\times & \left.
\int dt
d^2\u d^2\v \log^2 M_s^2(\u-\v)^2
G(u-x_1) G(v-x_1) G(x_3-u) G(x_4-v) \right\}
\label{tb6}
\eea
As in previous cases, the subdivergences have been regularized
and derivatives manipulated in order to make the integral over
internal variables as well as the Fourier transform respect to
external points well defined.
As in the case of diagram $i$, although each of these
diagrams depends on $M_s$, this dependence disappears when
we take into account the graphs obtained by
the permutation $A_0\leftrightarrow A_i$.
It is easy to prove that the scale derivative of
each of these graphs is proportional to a graph that
contains  graph $e$ as a subgraph and as a consequence
the scale derivative of these graphs is zero.

We conclude with the triple box (diagram r).
The contribution of this diagram to the amputated
4-point function is given by
\bea
\Gamma^r (x_i) &=&
{1 \over 4}
\left({e^2 \over m} \right)^4
{\partial \over \partial \x_2^i}
{\partial \over \partial \x_4^j}
T_i(x_2-x_1) T_j(x_4-x_3)
\nonumber \\
&\times &
\int_{u,v,y,z} G(u-x_1) G(y-u) G(x_3-y) T_k(v-u) T_l(z-y)
\nonumber \\
& &
\left \{ G(x_4-z)
\stackrel {\leftrightarrow} {{\partial \over \partial \z^l}}
G(z-v)
\stackrel {\leftrightarrow} {{\partial \over \partial \v^k}}
G(v-x_2) \right\}
\label{box1}
\eea
Because of the ``transversality'' of the gauge field propagator,
all derivatives inside the integral
can be rewritten as derivatives with respect to external points and
brought to the left to obtain the regulated form
\bea
\Gamma^r_{reg}(x_i) &=&
-\left({e^2 \over m} \right)^4
{\partial \over \partial \x_2^i}
{\partial \over \partial \x_4^j}
\left\{ T_i(x_2-x_1) T_j(x_4-x_3)
{\partial \over \partial \x_2^k}
{\partial \over \partial \x_4^l}
\right. \nonumber \\
&\times & \left.
\int_{u,v,y,z} G(u-x_1) G(y-u) G(x_3-y) T_k(v-u)  \right.
\nonumber \\
& & \left.
T_l(z-y) G(x_4-z) G(z-v) G(v-x_2) \right\}
\label{box2}
\eea
Away from coincident external points, one can see that
the integrals over internal points are now finite by power counting.
Furthermore,
the amplitude (\ref{box2}) has a well defined Fourier transform since
it contains two total derivatives with respect to external points.

\section{Renormalization Group Equations}
\setcounter{equation}{0}

In this section we shall show that the renormalized
4-point function satisfies the RGE equation and we
compute the $\beta$-function to three loops.
We recover the one loop result of \cite{bl}, namely
the zero of the $\beta$-function for the self-dual case,
and we find that the relation between the couplings
for which the $\beta$-function vanishes
is scheme dependent beyond one loop.
Scheme dependence arises from the two scale parameters
$M$ and $M_s$
introduced initially at one-loop order, and
renormalized amplitudes depend on the parameter
$\rho = M_s^2/M^2$.
As explained below (\ref{c12}), an additional scale parameter
in the regularization of graph n was fixed to simplify its
renormalized amplitude. This scale parameter
affects the RGE's beginning at four loops.

The RGE equation for
the 4-point correlation function is
\beq
\left[ M {\partial \over \partial M}
+ \beta {\partial \over \partial \lambda} \right]
\Gamma_{reg} = 0
\label{r1}
\eeq
Recall the anomalous dimension term is absent because there is no
field renormalization in this theory.
In order to check that (\ref{r1}) is satisfied,
we explicitly compute the scale derivatives of
the renormalized amplitudes obtained in the previous sections
and compare order by order in perturbation theory with
$\frac{\partial}{\partial \lambda}\Gamma_{reg}$.
The contribution to the $\beta$-function in each order can be uniquely
identified from the purely local term, in which
$\frac{\partial}{\partial\lambda}$
acts on the tree contribution (\ref{s1}).
Non-local contributions then provide an independent consistency check,
since they must appear with the correct coefficients to test
that subdivergences have been correctly handled by our method.

The $\beta$-function can be written
as a series expansion in the couplings $\lambda$, $e$
\beq
\beta(\lambda,e) = a_1 \lambda^2 + b_1 e^4 +
a_2 \lambda^3 + c_2 \lambda e^4 + \ldots  \ ,
\label{r2}
\eeq
where the subscript in the coefficients indicates the loop order.
In a theory with more than one coupling constant
the $\beta$-function is renormalization scheme independent
through one loop order but not beyond.
We will use capital letter superscripts to
denote that the amplitudes include now not only
the contribution of the particular diagram
drawn in Fig.~1 but also the contributions of graphs related by
Bose symmetry as well as time reflection.
All permutations of $A_0$, $A_i$ and $A_i^2$ vertices are also added.

We have already computed the one loop $\beta$-function in the purely
scalar subtheory
(see (\ref{s8})), while from (\ref{3.10})
one can immediately obtain
the scale derivative of the seagull graph
\bea
M {\partial \over \partial M} \Gamma^D_{reg} &=&
- { i e^4 \over \pi m \alpha^2}
\delta(x_1-x_2) \delta(x_1-x_3) \delta(x_1-x_4)  \nonumber \\
&=&
{ e^4 \over 4 \pi m \alpha^2  \lambda } \Gamma^A
\label{r3}
\eea
Thus
\bea
a_1 &=& {m \over \pi} \\
b_1 &=& - {1 \over 4 \pi m \alpha^2}
\label{r4}
\eea
and we see that
\beq
M {\partial \over \partial M} (\Gamma^C_{reg} + \Gamma^D_{reg}) = 0
\eeq
when
\beq
\lambda^2 =
{ e^4 \over 4 m^2 \alpha^2}
\label{sp}
\eeq
showing that scale invariance is recovered for the self-dual point
\cite{jp}.

At two-loop order
the $\beta$-function is renormalization scheme dependent,
as we shall see.
The scale derivative of the double bubble (\ref{4.1})
is trivially found to be
\beq
M {\partial \over \partial M} \Gamma^G_{reg} =
- { 2 m \lambda \over \pi} \Gamma^C_{reg}
\label{r5}
\eeq
The case of the ``ice cream cone'' diagram is
considerably more complicated.
Although it is true that the
regularized amplitude has a well defined scale derivative,
we must be able to recognize in the result the amplitude of
graphs that have already appeared at one loop order.
However, if we calculate the scale derivative of $f^h_{reg}$
from (\ref{4.9a}) we find
\bea
\sd f^h_{reg} (\x,\y,t) & = &
- \frac{1}{8}  (\nablar_\x - \nablar_\y)
\left[  \log M_s^2 (\x-\y)^2
(\nablaa_\x - \nablaa_\y) G(\x,t) G(\y,t) \right]
\nonumber \\
& & - \frac{1}{8}  \left[ (\nablar_\x +\nablar_\y)^2
+ 4im  \partial_t \right] G(\x,t) G(\y,t) \log M_s^2 (\x-\y)^2
\label{r6}
\eea
which is difficult to interpret immediately.
One can show that (\ref{r6}) and the regulated bubble (\ref{s3})
differ at most by a finite local term independent of $M$.
In order to determine this possible
$\beta$-function contribution,
we compute the Fourier transform of (\ref{r6})
(see appendix) and compare it with the
regularized expression for the bubble.
The result is
\beq
\sd f^h_{reg} (\x,\y,t;M_s)  =
2 \pi \delta(\x-\y)
\left[ G_{reg}^2(\x,t;M)
+ {i m \over 4 \pi} \log {\rho \over \rho_1}
\delta(t) \delta(\x)
\right]
\label{r8}
\eeq
where $\rho = M_s^2 / M^2$ and $\rho_1 = m \gamma / 4$.
Note that we used $M_s$ to regulate the subdivergence
of the graph,
but the scale derivative involves the one loop bubble.
The RGE requires that the renormalized bubble amplitude
appear with the assigned scale
parameter, {\it i.e.}, $M$.

Combining (\ref{r8}) with the ``ice cream cone'' amplitude
given by Eqs. (\ref{4.4}), (\ref{4.5}),
and adding the contribution of the
diagrams related by Bose symmetry and time inversion,
we obtain
\beq
M {\partial \over \partial M} \Gamma^H_{reg} (M_s)=
{ e^4 \over 2 \pi m \lambda \alpha^2 } \Gamma^C_{reg} (M)
+ { e^4 \over (2 \pi \alpha)^2 } \log {\rho_1 \over \rho} \Gamma^A
\label{r9}
\eeq
It is easy to check that (\ref{r5}) and (\ref{r9}) are consistent
with the one loop result and yield the two loop contributions:
\bea
a_2 &=& 0 \\
c_2 &=& {1 \over (2 \pi \alpha)^2} \log {\rho \over \rho_1}
\label{r10}
\eea
One can see explicitly that the coefficient $c_2$ depends on the
mass ratio we have chosen, $\rho = M^2_s / M^2$.
This means that the point for which the $\beta$-function vanishes
is also scheme dependent, and in general
there will be higher order corrections to the one-loop relation
(\ref{sp}).
However, in the renormalization scheme defined by
\beq
M^2_s = \rho_1 M^2 = {m \gamma \over 4} M^2
\label{r11}
\eeq
we have
$\rho = \rho_1$ and thus there is no contribution to the
two-loop $\beta$-function. As a consequence, in this scheme
the relation (\ref{sp})
between the couplings which gives a finite theory is
valid through two loops.

The scale derivatives of the three loop amplitudes
are found to be
\bea
M {\partial \over \partial M} \Gamma^K_{reg} &=&
- { 3 m \lambda \over \pi} \Gamma^G_{reg}  \\
M {\partial \over \partial M} \Gamma^L_{reg} (M_s,M) &=&
-{\lambda m \over \pi} \Gamma^H_{reg} (M_s)
+{ e^4 \over 2 \pi m \lambda \alpha^2} \Gamma^G_{reg} (M)
+ { e^4 \over (2 \pi \alpha)^2 } \log {\rho_1 \over \rho}
\Gamma^C_{reg} (M)
\\
M {\partial \over \partial M} \Gamma^M_{reg} &=&
{ e^4 \over 4 \pi m \lambda \alpha^2} \Gamma^H_{reg}   \\
M {\partial \over \partial M} \Gamma^N_{reg} (M_s) &=&
{ e^4 \over 4 \pi m \lambda \alpha^2} \Gamma^G_{reg} (M)
+ { e^4 \over (2 \pi \alpha)^2 } \log {\rho_1 \over \rho}
   \Gamma^C_{reg} (M)  \nonumber \\
& &
+ { \lambda e^4 m \over 4 \pi (2 \pi \alpha)^2 }
\log^2 {\rho_1 \over \rho} \Gamma^A
\label{r12}
\eea
where we have specified the mass parameter only when the mass scale
used in the regularization of the three loop
diagram is not the same as
the mass scale of the graph which appears in its scale derivative.
In general,
it is immediate to obtain these results from the regulated amplitudes
discussed in section V and only the scale derivative of diagram m
requires some further study.
As in the case of the ``ice cream cone'' (diagram h),
one can show by direct computation that
the difference between $M {\partial \over \partial M} \Gamma^m_{reg}$
(see (\ref{t7}))
and the regulated amplitude of diagram h
is purely local, finite and
independent of $M$. Fortunately, to determine this possible
local term we only
need to compute the Fourier transform
of a semi-local piece that does not appear in the
regulated form of the ``ice cream cone'', and we easily find that
there is no such local term.

When we substitute the scale derivatives
in the RGE (\ref{r1}),
we find the three-loop contributions to the $\beta$-function
\bea
a_3 &=& 0 \\
b_3 &=& 0 \\
c_3 &=& - {m \over 4 \pi (2 \pi \alpha)^2} \log^2 {\rho \over \rho_1}
\label{r13}
\eea
Note that if we choose
$M^2_s = \rho_1 M^2$
the three loop contribution to the $\beta$-function vanishes
and thus the theory is finite at the
self-dual point defined by (\ref{sp}).

To conclude we summarize the result for the $\beta$-function
\beq
\beta(\lambda,e) = {m \over \pi} \lambda^2
- {1 \over 4 \pi m \alpha^2} e^4
+ {1 \over (2 \pi \alpha)^2} \log {\rho \over \rho_1} \lambda e^4
- {m \over 16 \pi^3 \alpha^2} \log^2 {\rho \over \rho_1} \lambda^2 e^4
\label{beta}
\eeq

\section*{Acknowledgement}

We thank G. Dunne, A. Lerda and G. Zemba for useful discussions.
G.~Lozano thanks R.~Jackiw and J.~Negele for hospitality at
the Center for Theoretical Physics.
N. Rius acknowledges the Ministerio de Educaci\'on y Ciencia (Spain)
for a Fulbright/MEC Scholarship.

\appendix
\section*{}
\setcounter{equation}{0}

In this appendix we shall calculate the scale derivative
of the ice cream graph h.
In order to do so, we shall calculate its Fourier transform
and show how it is related to the bubble graph c.
Let us start by recalling that
\bea
R(\x,\y,t) = \sd f^h_{reg} &=&
- \frac{1}{8}  (\nablar_\x - \nablar_\y)
\left[  \log M_s^2 (\x-\y)^2
(\nablaa_\x - \nablaa_\y) G(\x,t) G(\y,t) \right]
\nonumber \\
& & - \frac{1}{8}  \left[ (\nablar_\x +\nablar_\y)^2
+ 4im  \partial_t \right] G(\x,t) G(\y,t) \log M_s^2 (\x-\y)^2
\label{a1}
\eea
It is convenient to define variables $\u=(\x+\y)/2$ and
$\v=(\x-\y)/2$.
After manipulations of derivatives and use of (\ref{1.4}),
the Fourier transform of $R$ can be shown to be
\beq
\hat R(\p_1,\p_2,\omega)=
4 \{ i\p_2 \cdot \f_3 + [-\p_1^2-\p_2^2 + 4m\omega]f_2 \}
\eeq
where
$\p_1 = \p_x + \p_y$, $\p_2 = \p_x-\p_y$, and
\bea
f_2&=&-\frac{1}{8}\int d^2\u d^2\v dt \log 4M_s^2\v^2
G(\sqrt{2}\u,t) G(\sqrt{2}\v,t)
e^{-i\p_1 \cdot \u -i\p_2 \cdot \v +i\omega t} \\
\f_3&=&-\frac{1}{4}\int d^2\u d^2\v dt (\nabla_v\log4M_s^2\v^2)
G(\sqrt{2}\u,t) G(\sqrt{2}\v,t)
e^{-i\p_1 \cdot \u -i\p_2 \cdot \v +i\omega t}
\eea

The calculation of $\f_3$ is straightforward.
A conventional non-relativistic loop integral gives the result:
\beq
\p_2 \cdot \f_3 =-\frac{m}{4}\log \left( \frac
{- 4m\omega + \p_1^2}{- 4m\omega + \p_1^2 + \p_2^2}\right)
\eeq

The loop integral representation of $f_2$ is infrared divergent because
it contains the factor $1/q^2$ which is the Fourier transform
of the logarithmic term. Instead we
calculate $f_2$ by the following 3-step technique:

i)a trivial calculation of the transform of
$G(\sqrt{2}\u,t)$ with respect to $\u$ and $t$,

ii)a not-so-trivial calculation of the transform of
$\log 4M_s^2\v^2 G(\sqrt{2}\v,t)$
with respect to $\v$ and $t$,

iii)a final convolution integral over $\omega$ to obtain the net result.

Step ii) is done by considering the function
$(\v^2)^\alpha G(\sqrt{2}\v,t)$,
calculating its Fourier transform and taking the derivative with respect
to $\alpha$ at $\alpha = 0$ to obtain the logarithmic factor.
Several special function formulae \cite{gr} are required for this task.
The final result is
\beq
f_2 =\frac{im}{8[-\p_1^2-\p_2^2 +4m\omega]}
\log \left[ \left({16 M_s^2 \over \gamma^2}\right)
\frac{- 4m\omega + \p_1^2}
{(-4m\omega +\p_1^2 + \p_2^2)^2}\right]
\eeq

Then,
\beq
\hat{R}(\p_1,\p_2,t)= - \frac{im}{2}\log\left[
\left(-\omega+ {\p_1^2 \over 4m}\right)
\frac {m\gamma^2} {4 M_s^2} \right]
\eeq
Now, using (\ref{s2}) and comparing with the regulated amplitude of
the bubble in momentum space (\ref{s5}) we get the relation
\beq
\hat R(\p_x,\p_y,t)=2\pi
\hat G^2 \left(\p_x+\p_y, \omega; M
\sqrt{\frac{\rho}{\rho_1}}  \right)
\eeq
where $\rho_1 = m \gamma /4$
and $\rho = M_s^2 / M^2$, which is the Fourier transform of
(\ref{r8}).

One should note that the regularized amplitude of graph h,
see (\ref{4.9a}), differs from its scale derivative only by the
exponent of the logarithmic factor. The loop integral representation
of its Fourier transform is also infrared divergent.
However the transform is well defined, and a technique similar to that
described above is required to calculate it.

\begin{figure}
\caption
{Figure 1.  Diagrams which contribute to the 4-point function $\Gamma(x_1,
x_2, x_3, x_4)$ through three-loop order.  Dots represent the $A_i$ vertex
with derivative coupling.  Diagrams related by permutations must be added as
explained in section III.}
\end{figure}

\end{document}